\documentclass[10pt,a4paper]{article}
\usepackage[intlimits]{amsmath}
\usepackage[psamsfonts]{amssymb}
\usepackage{layout}
\usepackage{enumerate}
\usepackage{longtable}
\usepackage{array}

 \allowdisplaybreaks
 \allowdisplaybreaks
 \allowdisplaybreaks
 \allowdisplaybreaks 

 \allowdisplaybreaks

\begin{document}

Journal of Russian Laser Research, Volume 22, Number 5, 2001, p. 475-479.

\begin{center}
{\Large \textbf{HIDDEN VARIABLES AND QUANTUM STATISTICS NATURE}\\[0pt]
\vspace{1cm} }

{\Large \textit{T.F. Kamalov}\\[0pt]
}

{\Large \vspace{0.5cm} Physics Department\\[0pt]
Moscow State Opened University\\[0pt]
107996 Moscow, 22, P. Korchagin str., Russia\\[0pt]
E-mail: ykamalov@rambler.ru\\[0pt]
}
\end{center}
\begin{abstract}
It is shown that the nature of quantum statistics can be clarified
by assuming the existence of a background of random gravitational
fields and waves, distributed isotropically in the space. This
background is responsible for correlating phases of oscillations
of identical microobjects. If such a background of random
gravitational fields and waves is considered as hidden variables
then taking it into account leads to the Bell-type inequalities
that are fairly consistent with the experimental data.
\end{abstract}

\bigskip

Quantum theory is a statistical theory, which at the same time does not lend
itself to investigation of its statistics nature, this problem being
considered as being beyond its scope. Quantum theory does not deal with the
causes of quantum phenomena; it postulates the classically inexplicable
phenomena of a quantum microcosm observed in experiments as its axioms. Such
an approach, although not introducing errors, does not explain the
experimentally observable phenomena, leaving them incomprehensible from the
classical viewpoint and giving rise to all sorts of paradoxes. Quantum
theory lacks the classical logic and the classical causality, hence the
classical axiomatics, which makes this theory, from the classical physics
viewpoint, rely on the method of indirect computations.

Are classical causality and classical logic absent in quantum theory only or
in nature as well? The absence of classical causality and classical logic in
the theory does not imply their absence in the nature.

Now, let us try to single out the basic classically incomprehensible
concepts of quantum theory. First, it is the wave-particle dualism. Taking
into account all the above-mentioned, a particle could acquire wave
properties, being influenced by a wave background. Second, it is
Heisenberg's uncertainty principle. Due to the influence of nonremovable
background on a measurement, it is impossible to measure the values
precisely. Third, it is the energy balance in an atom. From the classical
physics viewpoint, an electron moving in the electric field of the nucleus
should emit electromagnetic radiation. Can we assume that the background of
the whole spectrum of frequencies gives energy to the electron, the latter
re-emitting it, and that the energy balance in the atom could then be
maintained?

We can complete quantum theory with hidden variables without altering the
mathematical apparatus of quantum mechanics. Does a comprehensible theory
result?

The issue of the necessity to complete the quantum theory was first
considered in the study by A. Einstein, B. Podolsky and N. Rosen
(hereinafter, EPR) [1]. Let us consider the EPR effect. Two particles, $A$
and $B$, at the initial moment interact and then scatter in opposite
directions. Let the first of them be described by the wavefunction $\psi
_{A} $, the other one by $\psi _{B}$. The system of the two particles $A$
and $B$ is described by the wavefunction $\psi _{AB}$. With this,

\begin{center}
$\psi _{A}\neq \psi _{AB}$, $\psi _{B}\neq \psi _{AB}$, $\psi _{AB}\neq \psi
_{A}\psi _{B}$, or, $P_{AB}\neq P_{A}P_{B}$.
\end{center}

For independent events $P_{A}$ and $P_{B}$, according to probability theory,

\begin{center}
$P_{AB}=P_{A}P_{B}$.
\end{center}

Where could the dependence of the object $A$ on the object $B$ and vice
versa originate from, these objects $A$ and $B$ being considered as distant
and noninteracting? The EPR authors arrived at the conclusion on the
incompleteness of the quantum--mechanical description. To solve this
contradiction, an idea has been put forward in [1] on the existence of
hidden variables that would make it possible to interpret consistently the
results of the experiments without altering the mathematical apparatus of
quantum mechanics.

Later, it has proved by von Neumann [2] that quantum mechanics axiomatics
does not allow the introduction of hidden variables. It is, however,
important that the argument presented in [2] is not valid in certain cases,
e. g., for pairwise observable microobjects (for Hilbert space with pairwise
commutable operators) [3]. In 1964, J. S. Bell [4] formulated an
experimental criterion enabling to decide, within the framework of the
problem statement [1], on the existence of hidden variables. The essence of
the experiment proposed by Bell is as follows.

Let us consider the experimental scheme of EPR. Let there be two photons
that can have orthogonal polarizations $A$ and $B$ or $A^{^{\prime }}$ and $%
B^{^{\prime }}$, respectively. Let us denote the probability of observing a
pair of photons with polarizations $P$ and $Q$ as $\psi _{PQ}^{2}$. Bell
introduced the quantity

\begin{center}
$\left\vert \left\langle S\right\rangle \right\vert =\frac{1}{2}\left\vert
\psi _{AB}^{2}+\psi _{A^{^{\prime }}B}^{2}+\psi _{AB^{^{\prime }}}^{2}-\psi
_{A^{^{\prime }}B^{^{\prime }}}^{2}\right\vert $,
\end{center}

called the Bell's observable; it has been shown that if hidden variables do
exist, then

\begin{center}
$\left\vert \left\langle S\right\rangle \right\vert \leq 1$.
\end{center}

The possibility of experimental verification of the actual existence of
hidden variables has been demonstrated in [4]. The above inequality are
called Bell's inequalities. A series of experiments has shown that there is
no experimental evidence of the existence of hidden variables as yet, and
the existing theories comprising hidden variables are indistinguishable
experimentally. In quantum theories with hidden variables, the wavefunction

\begin{center}
$\psi =\psi \left( \lambda _{i}\right) $
\end{center}

is a function of hidden variables $\lambda _{i}$.

Let us consider a physical model with gravity background (i. e., the
background of gravity fields and waves) playing the role of hidden variables
[5-8].

This is only one of many possible versions. We could consider as hidden
variables, for example, the electromagnetic background. We shall not discuss
here the reasons for this version being unfounded, and we shall not consider
it in the present study.

So, let us regard the gravitational background as hidden variables. The
gravitational background could be considered negligible and not affecting
the behavior of quantum microobjects. Let us verify whether this is correct.
The quantitative assessments of the gravitational background influence on
the quantum microobjects' behavior have not been performed due to the former
having never been examined. The quantum effects are small as well, but their
quantitative limits are known and are determined by the Heisenberg
inequalitie. Let us demonstrate the gravitational background being random
and isotropic to affect the phases of microobjects separated in the space
and not interacting. Then we can calculate the correlation factor for these
microobjects, hence, the Bell's observable $S$. Having determined the upper
limit for $S$, we shall get the refined Bell's inequalities taking into
consideration the influence of the gravitational background. Comparing these
with the experimental data for the Bell's observable, we can verify the
correctness of our approach.

By now, hundreds of experimental studies have been performed on measurement
of the Bell's observable. It can be positively stated that the experimental
value of the Bell's observable has been determined to comply with the
expression $\left| \left\langle S\right\rangle \right| \leq \sqrt{2}$.

Relative oscillations $\ell ^{i}$, $i=0,1,2,3$ of two particles in gravity
fields are described by the deviation equations. In this particular case,
the deviation equations are converted into the oscillation equations for two
particles:
\begin{equation*}
\overset{..}{\ell }^{1}+c^{2}R_{010}^{1}\ell ^{1}=0,\quad \omega =c\sqrt{%
R_{010}^{1}}.
\end{equation*}

It should be noted that relative oscillations of micro objects $A$ and $B$
do not depend on the masses of these, but rather on the Riemann tensor of
the gravity field. This is important, since in the microcosm we are deal
with small masses. Taking into account the gravity background, the
microobjects $A$ and $B$ shall be correlated. It is essential that in
compliance with the gravity theory, the deviation equation only make sense
for two objects, and it is senseless to consider a single object. Therefore,
the gravity background complements the quantum--mechanical description and
plays the role of hidden variables. On the other hand, the von Neumann
theorem on impossibility of introduction hidden variables into quantum
mechanics is not applicable for pairwise commuting quantities (Gudder's
theorem [3]). The introduction of hidden variables in the space with
pairwise commuting operators is appropriate.

The solution of the above equation has the form

\begin{center}
$\ell ^{1}=\ell _{0}\exp (k_{a}x^{a}+i\omega t)$, $a=1,2,3$,
\end{center}

were we assume the gravity background to have a random nature and to be
described, similarly to quantum--mechanical quantities, with probabilistic
observations. Each gravity field or wave with the index $n$ and Riemann
tensor $R(n)$ and random phase

\begin{center}
$\Phi (n)=\omega (n)t=c\sqrt{R_{0101}^{1}(n)}t$,
\end{center}

should be matched by a quantity $\ell ^{i}(n)$. Therefore, taking into
account the gravity background, i. e. the background of gravity fields and
waves, the particles take on properties described by $\ell ^{i}(n)$.

In the present study, we consider only the gravity fields and waves, which
are so small that alter the variables of microobjects $\Delta x$ and $\Delta
p$ beyond the Heisenberg inequality $\Delta x\Delta p\geq h$. Strong fields
are adequately described by the classical gravity theory, so we do not
consider them here. Let us emphasize that the assumption on existence of
such a negligibly small background is quite natural. With this, we assume
the gravity background to be isotropically distributed over the space.

Regarding the quantum microobjects in the curved space, we must take into
account the scalar product $g_{\mu \nu }A^{\mu }B^{\nu }$ of two
4-dimensional vectors $A^{\mu }$ and $B^{\nu }$, where for weak
gravitational fields it is possible to employ the value $h_{\mu \nu }$,
which is the solution of Einstein equations for the case of weak
gravitational field in harmonic coordinates and having the form:
\begin{equation*}
\begin{array}{c}
h_{\mu \nu }=e_{\mu \nu }\exp (ik_{\gamma }x^{\gamma })+\qquad \\[2mm]
\qquad +e^{\ast }\exp (-ik_{\gamma }x^{\gamma }),%
\end{array}%
\end{equation*}%
\begin{equation*}
\ g_{\mu \nu }=\delta _{\mu \nu }+h_{\mu \nu },
\end{equation*}%
where the value $h_{\mu \nu }$ is called the metrics disturbance, and $%
e_{\mu \nu }$, the polarization. Therefore, we shall consider the hidden
variables $h_{\mu \nu }$ as being the disturbances of the metrics as
distributed in the space with the yet unknown distribution function $\rho
=\rho (h_{\mu \nu })$. Hereinafter, the indices $\mu ,\nu ,\gamma $ possess
values 0,1, 2, 3.

Then the coefficient of correlation $M$ of projections of unit vector$\
\lambda ^{i}$ of the hidden variables onto directions $a^{k}$ and $b^{n}$
specified by the polarizers is
\begin{equation*}
M=\left\langle AB\right\rangle =\langle \lambda ^{i}a^{k}g_{ik}\lambda
^{m}b^{n}g_{mn}\rangle
\end{equation*}

were $i,k,m,n$ possess 0,1,2,3 and

\begin{center}
$\theta =\left( \overrightarrow{a}^{\wedge }\overrightarrow{b}\right) $, $%
\alpha =\left( \overrightarrow{\lambda }^{\wedge }\overrightarrow{a}\right) $%
, $\beta =\left( \overrightarrow{\lambda }^{\wedge }\overrightarrow{b}%
\right) $,
\end{center}

and thus

\begin{center}
$M=\frac{1}{\pi }\int_{0}^{2\pi }d\alpha \cos \alpha \cos \left( \alpha
+\theta \right) =\cos \theta $.
\end{center}

Then, for $\theta =\frac{\pi }{4}$, we obtain the maximum value of the
Bell's observable $S$

\begin{center}
$\langle S\rangle =\frac{1}{2}[\langle AB\rangle +\left\langle A^{^{\prime
}}B\right\rangle +\left\langle AB^{^{\prime }}\right\rangle -\left\langle
A^{^{\prime }}B^{^{\prime }}\right\rangle ]=$

$=\frac{1}{2}[\cos (-\frac{\pi }{4})+\cos (\frac{\pi }{4})+\cos (\frac{\pi }{%
4})-\cos (\frac{3\pi }{4})]=\sqrt{2}$,
\end{center}

which agrees fairly with the experimental data.

The Bell-type inequality in our assumptions ( in view of taking into account
the gravitational background) should have the form

\begin{center}
${\left\vert \left\langle S\right\rangle \right\vert \leq \sqrt{2}}$.
\end{center}

Therefore, we have shown that the classical physics with the gravitational
background gives a value of the Bell's observable that matches both the
experimental data and the quantum mechanical value of the Bell's observable.
To sum up, the description of microobjects by the classical physics
accounting for the effects brought about by the gravitational background is
equivalent to the quantum-mechanical descriptions, both agreeing with the
experimental data.

From the experiment viewpoint, both of these descriptions are equivalent;
however, employing the quantum-mechanical descriptions demands using the
quantum mechanical axioms. In addition, plausible arguments should be given
that these predictions and interpretation are experimentally distingvishable
from existing knowlege.

\end{document}